# Astro2020 Science White Paper

# An Interdisciplinary Perspective on Elements in Astrobiology: From Stars to Planets to Life

**Thematic Areas:** ☒ Planetary Systems  ☒ Star and Planet Formation
☐ Formation and Evolution of Compact Objects  ☐ Cosmology and Fundamental Physics
☒ Stars and Stellar Evolution  ☐ Resolved Stellar Populations and their Environments
☐ Galaxy Evolution  ☐ Multi-Messenger Astronomy and Astrophysics

**Principal Author:**
Name: Natalie Hinkel
Institution: Southwest Research Institute
Email: natalie.hinkel@gmail.com

**Co-Authors :** Hilairy Hartnett [Arizona State University], Carey Lisse [Johns Hopkins University Applied Physics Laboratory], Patrick Young [Arizona State University]

**Co-signers:**

**Abstract**
Stellar elemental abundances direct impact planetary interior structure and mineralogy, surface composition, and life. However, the different communities that are necessary for planetary habitability exploration (astrophysics, planetary science, geology, and biology) emphasize different elements. If we are to make progress, especially in view of upcoming NASA missions, we must collectively broaden our communication regarding lists of useful elements, their impact on planetary systems, and how they can be observed in the near and long term.

**Elements Across Disciplines**
The discovery of planets outside of our Solar System has created, not only a new field of exploration, but also a bridge between many interdisciplinary sub-fields with little-to-no history of communication. For example, when measuring element abundances in stars, spectroscopists often focus on elements with multiple, strong absorption lines in the wavelength band that will ensure better accuracy for their abundance determinations. Spectroscopists also tend to measure elements that cover a variety of nucleosynethetic origins, such as alpha-type, iron-peak, or neutron capture, since the different processes that create these elements are indicative of historical events prior to the star's formation.

Geophysicists and mineral physicists often concentrate on processes that occur on Earth's surface and within its interior. Predominantly, this entails how gravity, heat flow, seismic waves, radioactivity, fluid dynamics, and magnetism have shaped the planet − influences that essentially ignore life. When considering exoplanets, it is possible to measure the stellar elemental abundances most important for planetary and rock formation -- such as Fe, Mg, Si, C, O, Al, Ca and to a lesser extent C, Na, P, S, Ni, then extrapolate the likely planetary interior structure and mineralogy using mass-radius models. However, when using stellar abundances to understand planetary interiors, the standard model is made complicated because it does not matter whether an element is considered "volatile" or "refractory" from an astrophysical standpoint; it only matters what reactions these elements participate in and how they are redistributed in the planet. In addition, there is a scarcity of key elements from stellar spectroscopy that are necessary for planetary formation and evolution.

However, biology and the fundamental mechanisms of living systems, extending from more standard surface conditions to extremophiles, require a different suite of major (H, C, N, O, P, S), intermediate (F, Na, Mg, Cl, K, Ca, Mn, Fe), and minor (V, Cr, Co, Ni, Cu, Zn, Mo) elements. From the viewpoint of exoplanets and habitability, we must also consider that not all life, or even most life, exists on the surface of a planet.

**Stellar Abundances of Planet Hosts**
The iron ratio within stars, or [Fe/H], is one of the most important properties of a star: it indicates age and history, has hundreds of lines within a star's optical spectrum, and is often used as a proxy for overall metallicity, or those elements that are heavier than H or He. In Fischer & Valenti (2005), the "planet-metallicity" relationship was introduced, indicating that stars that host giant, gaseous planets are preferentially enriched in heavy elements compared with stars that do not host planets. An example of abundance exploration in stellar hosts is the Near InfraRed

Disk Survey (NIRDS, Lisse et al. 2012) team who observed the absorption lines of 16 Kepler host stars selected by Kane et al. (2016) as most likely to harbor habitable Earth-like planets. Three of the stars show unusual Fe/Si vs Mg/Si atomic abundance ratios, suggesting their Earth-like planets might have very different sized cores and mantles than our own. Another has a very low C/Si ratio, suggesting that organic material needed for life might be rare on any terrestrial planets in its system. As a check, they performed the same study on 14 bright nearby G-star planet-hosting systems the NIRDS database, and found a similar occurrence rate of unusual abundances and improper stellar characterizations. However, while dozens of different studies have confirmed that giant planet-hosting stars are enriched in [Fe/H], specifically, the same trend has not been found when looking at other individual elements, such as C, Mg, Si, or Ni.

Instead of relying on statistical techniques that analyze the differences between known and unknown host stars one element at a time, Hinkel et al. (2019) employ a machine learning algorithm that allows multiple elements to be compared at the same time, as an ensemble. They are therefore able to define markers within the chemistry of the host star, as opposed to more standard planetary detection techniques which rely on physical stellar properties, that may indicate the presence of a yet-undetected giant planet. In order to take advantage of a variety of element abundances measured within nearby stars, they use the Hypatia Catalog of stellar abundances (Hinkel et al. 2014, www.hypatiacatalog.com, shown in Fig 1.).

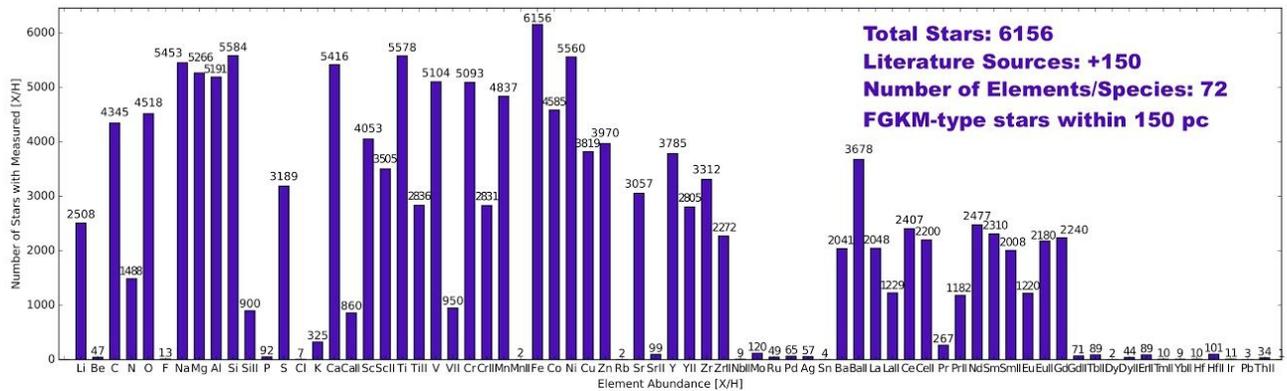

**Fig 1.** The number of stars for which particular elements have been measured as per the Hypatia Catalog, i.e. the largest catalog of elemental abundances for stars near to the Sun (www.hypatiacatalog.com, Hinkel et al. 2014). Of note are the lack of measurements for biologically important, "interdisciplinary elements" such as N, F, P, Cl, and K.

Hinkel et al. (2019) utilized their machine learning algorithm to predict the likelihood that +4200 FGK-type stars host a giant exoplanet, implementing five different ensembles of elements composed of volatiles, lithophiles, siderophiles, and Fe. Between the ensembles they found that C, O, and Fe, as well as Na although to a lesser extent, are the most important features for predicting giant exoplanet host stars. When they allowed the algorithm to predict on a hidden

"golden set" of *known* host stars, they had an average of ~75% probability of hosting a giant exoplanet, where more than half had a prediction probability ~90%. In addition, they investigated archival HARPS radial velocities for the top 30 predicted planet-hosting stars and found significant trends that HIP 62345, HIP 71803, and HIP 10278 host long-period giant planet companions with estimated minimum $M_p \sin i$ of 3.7, 6.8, and 8.5 $M_J$, respectively. They conclude that those stars with a high prediction probability are therefore likely to host a giant planet.

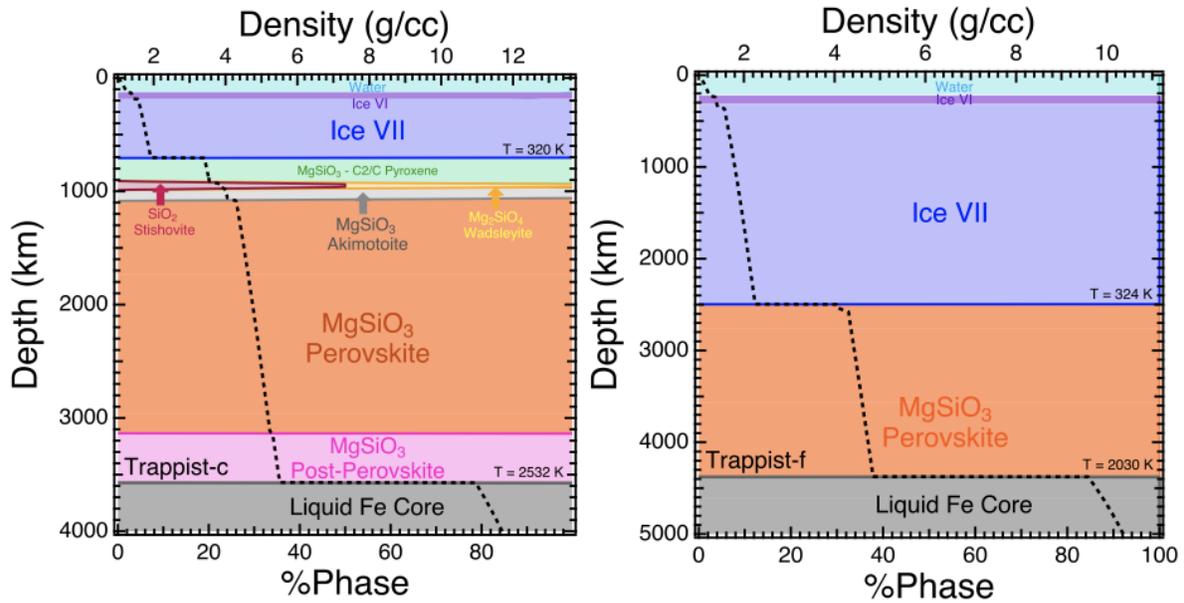

**Fig 2:** Phase diagram versus depth of the TRAPPIST-1c planet (left) and the TRAPPIST-1f planet (right), from Unterborn et al. (2018), as calculated by the ExoPlex mass-radius-composition calculator for the best-fit interiors using stellar abundances from the Hypatia Catalog (Hinkel et al. 2014).

**Stellar Hosts to Planetary Interiors**
Stars and planets are formed at the same time from the same bulk materials within the molecular cloud. Since it is not currently possible to directly measure the composition or surface materials of a planet, we must use stellar elemental abundances as a proxy for the interior makeup of the planet. For example, Thiabaud et al. (2015) analyzed whether the C, O, Mg, Si, and Fe were the same within the planet as a star, using a planet formation and composition model, and determined that these important rock-forming elements were the same in both the star and planet. To this end, a variety of models have utilized stellar abundances to constrain the interiors of terrestrial exoplanets, i.e. ExoPlex -- which determines the mineralogy and density of planetary interiors (Unterborn et al. 2016, 2018a/b, Hinkel & Unterborn 2018) and a Bayesian-based probabilistic inverse analysis (Dorn et al. 2015,2017). For example, stellar abundances from the Hypatia Catalog (Hinkel et al. 2014) were used to determine the interior structure of the TRAPPIST-1 planets (Gillon et al. 2017) via ExoPlex (Unterborn et al. 2018). Figure 2 shows the phase diagram of TRAPPIST-1c, with 8wt% water, versus -1f, with 50wt% water. The Earth, by comparison, is 0.02wt% water, meaning that these planets are not only water worlds, but likely do not have the important geochemical or elemental cycles that are absolutely necessary for life. The overall, holistic habitability of an exoplanet is dependent on its surface conditions, internal structure, mineralogy, and atmosphere (Foley & Driscoll 2016), where we must use host star measurements to understand the majority of these properties.

**Interdisciplinary Elements**

The discovery and subsequent drive to characterize nearby exoplanets has laid the foundation between stellar astrophysics, planetary science, geology, and biology. As a result of these new interdisciplinary relationships, it has become clear that there is a dearth of data for many of these planetary and biologically important elements. The lack of measurements for these interdisciplinary elements, such as N, F, P, Cl, and K, are shown in Fig. 1, which is a histogram produced from the Hypatia Catalog, the largest database of stellar abundances for stars near to the Sun (Hinkel et al. 2014). The Hypatia Catalog is a comprehensive collection of literature data and encompasses the largest number of element abundances of any dataset or survey.

Table 1: A list of elements commonly measured in stars (blue) per the Hypatia Catalog (Hinkel et al. 2014) and some of those important for planet formation (red) and biology (green). Note, elements labeled "Uncommon in Stars" compare those important to rocks/planets and life versus the number of stars for which those elements have been measured within stars.

**Major Stars:** Na, Mg, Al, Si, Ca, Ti, V, Cr, Mn, Fe, Cu
**Intermediate Stars:** C, O, Sc, Cu, Zn, Y, Zr, Ba II
*Uncommon in Stars: P, N, F, Cl, K*
**Major Rock:** O, Mg, Al, Si, K, Ca, Fe
**Minor Rock:** C, Na, P, S, Ni
**Major Life:** H, C, N, O, P, S,
**Intermediate Life:** F, Na, Mg, Cl, K, Ca, Mn, Fe
**Minor Life:** V, Cr, Co, Ni, Cu, Zn, Mo

When going from stars to exoplanets -- geophysics, and geobiology are the next step after planet formation. We compare those elements most influential to stellar spectroscopy, mineral physics, and astrobiology (shown in Table 1), the physical and chemical processes that govern their important interactions, as well as the flow of information between disciplines. The details of these interdisciplinary fields need to be made accessible, such that observations and trends about elements that have been here-to-fore difficult to measure or underappreciated can be preferentially targeted ("Uncommon in Stars") in future observations and missions. Our ultimate goal is to understand and define holistic planetary habitability from star to rock to microbe.

Looking ahead, we will also need to understand how a protoplanetary disk fractionates especially as it pertains to the distance from the star, since condensation temperatures of various elements will ultimately change how the properties, and where they end up, within the disk. Multiplanetary systems also have a big influence on the elements available to comprise a planetary core and its outer layers. In addition, planetary heating sources need to be considered, especially radiogenic heating which is produced from the decay of radioactive elements in the interior. The amount of heat produced by the radioactive decay of the radionuclides $^{235}$U, $^{238}$U, $^{232}$Th, and $^{40}$K is entirely dependent on the absolute abundances of these isotopes inherited from the host star upon formation. The radiogenic heat sources account for 30%–50% of the Earth's current heat budget and were present in larger fractions 4.5 billion years ago at the Earth's birth. (Schubert et al., 1980; Huang et al., 2013). Of the three radiogenic heat producing elements, Th and U are both refractory, such that their stellar abundance ratio relative to other refractory elements (e.g. Si) are likely mirrored in the resulting rocky planets (Unterborn et al., 2015).

**Future Observations**
With the discovery of ~4000 exoplanets[1] with a variety of masses, radii, compositions, and geometric configurations from their host star (and other planets), these sub-disciplines have found themselves asking important questions and looking to each other for answers: What are the elements most important to rocky planetary formation and to biology? and how much does an average star have? What is the cut-off between a rocky (Earth-like) and gas giant (Jupiter-like) planet? What does the interior of a planet 2.5 times the radius of the Earth look like? What is the compositional difference between planets that form far away from the host star compared with those that form nearby? or is Earth H- and N- poor because of Jupiter? Is elemental distribution enough to model whether a planet is potentially habitable? What is the minimum amount of bio-essential elements (e.g., P) that could sustain life on a exoplanet? What is the most compositionally extreme planet (compared to the Solar System planets) that can be formed? Does life create predictable changes in elemental cycling, such that we can potentially detect these elements and how do geophysical processes influence these changes? We do not know the answers to these questions. But they are all important to the definition and characterization of habitability, and will be at the forefront of exploration with the launch of large upcoming missions such as NASA's JWST and WFIRST.

---

[1] https://exoplanetarchive.ipac.caltech.edu/